\documentclass[superscriptaddress,prb,aps,twocolumn]{revtex4-1}
\usepackage{dcolumn}
\usepackage{bm}

\usepackage[colorlinks,hyperindex, urlcolor=blue, linkcolor=blue,citecolor=black, linkbordercolor={.7 .8 .8}]{hyperref}
\usepackage{graphicx}
\usepackage{amsfonts}
\usepackage{amsmath}
\usepackage{amssymb}
\usepackage{amsbsy}
\usepackage{nicefrac}
\newcommand{\iv}{$I$-$V$}

\begin{document}

\title {Evidence for defect-mediated tunneling in hexagonal boron nitride-based junctions}

\author{U. Chandni}
\email{chandniu@caltech.edu}
\address{Institute for Quantum Information and Matter, Department of Physics, California Institute of Technology, 1200 E. California Blvd., Pasadena, California 91125, USA}
\author{K. Watanabe}
\address{National Institute for Materials Science, 1-1 Namiki Tsukuba, Ibaraki 305-0044, Japan}
 \author{T. Taniguchi}
\address{National Institute for Materials Science, 1-1 Namiki Tsukuba, Ibaraki 305-0044, Japan}
\author{J. P. Eisenstein}
\address{Institute for Quantum Information and Matter, Department of Physics, California Institute of Technology, 1200 E. California Blvd., Pasadena, California 91125, USA}

\begin{abstract}
We investigate tunneling in metal-insulator-metal junctions employing few atomic layers of hexagonal boron nitride (hBN) as the insulating barrier. While the low-bias tunnel resistance increases nearly exponentially with barrier thickness, subtle features are seen in the current-voltage curves, indicating marked influence of the intrinsic defects present in the hBN insulator on the tunneling transport. In particular, single electron charging events are observed, which are more evident in thicker-barrier devices where direct tunneling is substantially low. Furthermore, we find that annealing the devices modifies the defect states and hence the tunneling signatures.
\end{abstract}

\date{\today}

\maketitle

Van der Waals heterostructures, where layered stacks of two dimensional materials are embedded in precisely desired patterns, have gained immense attention in recent times \cite{geim13,wang13}. Such tailor-made structures of graphene, hexagonal boron nitride (hBN), metal dichalcogenides and other 2D materials offer exotic device geometries to explore new physics. Tunnel junctions with an atomically thin insulator barrier sandwiched between atomic layers of 2D materials form an interesting structure in this respect. In conventional two dimensional semiconductor double layer structures, tunneling has shown remarkable features, including resonant tunneling, Coulomb correlations at high magnetic fields and Landau-level spectroscopy \cite{smoliner96,eisenstein91,eisenstein92,brown94}. In the regime of 2D layered materials, hBN with a band gap of $\sim$~5.9~eV~\cite{watanabe04} is an ideal candidate for an insulating barrier~\cite{lee11}. Recent studies on heterostructures with single and bilayer graphene as electrodes and hBN as the insulator have shown interesting features, including a very strong negative differential resistance (NDR)~\cite{britnell12,britnell13,mischchenko14,fallahazad14}.  In all of these transport studies hBN is assumed to be a benign element. However, from a materials perspective there have been extensive efforts to understand the underlying structure and defect mechanisms in thin layers of hBN, which can have important consequences on electrical transport\cite{Gibb13,cretu14,Rasool15}. Here we present detailed tunneling transport measurements on simple junctions consisting of metal or graphite electrodes separated by a thin hBN layer.  Our results demonstrate that the hBN insulator can yield unexpected transport signatures suggestive of defect-mediated tunneling processes.

\begin{figure}
\centering
\includegraphics[scale=0.45]{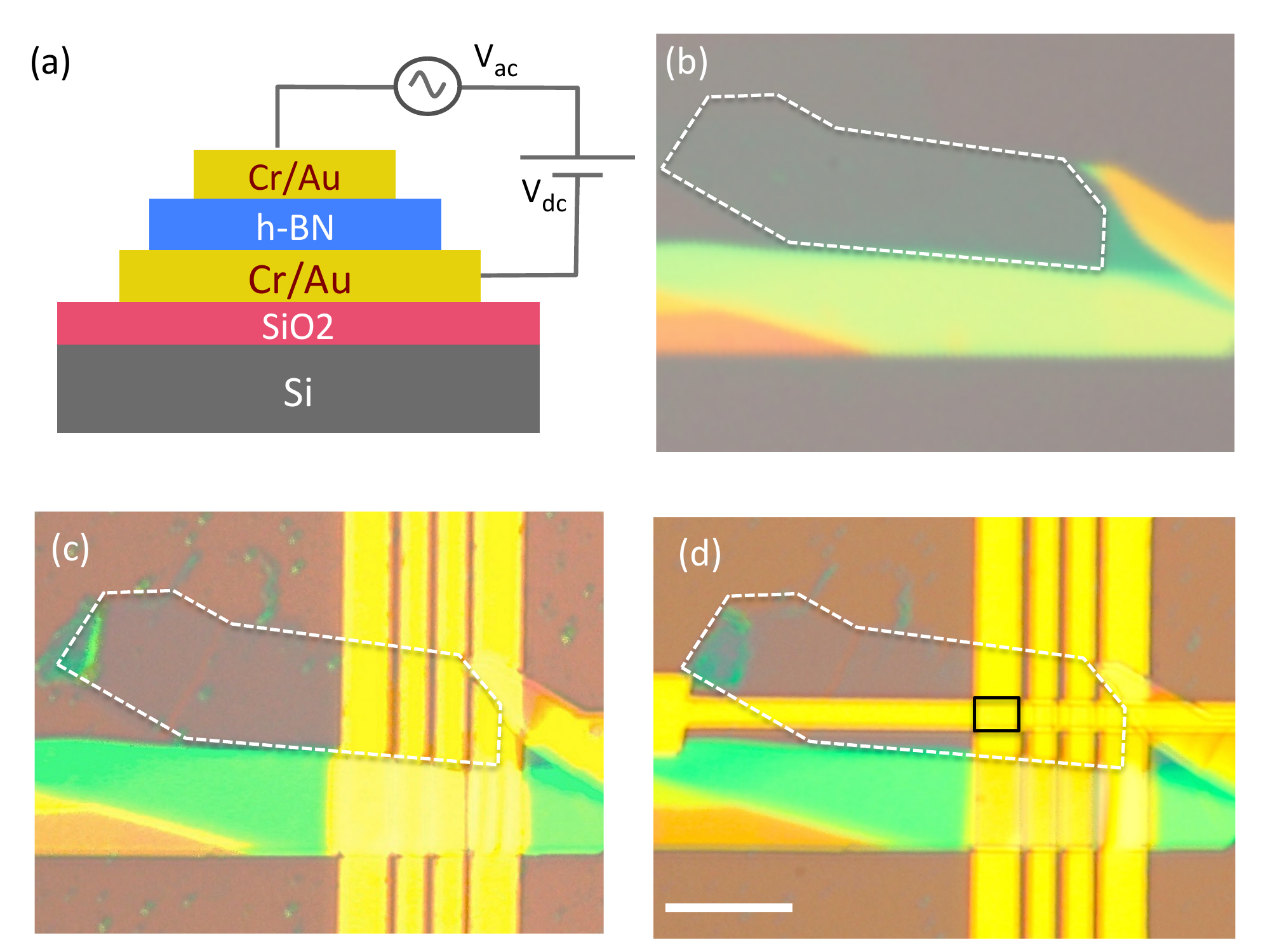}
\caption{(a) Device structure for h-BN tunnel devices with Cr/Au electrodes. (b) An optical micrograph of a thin hBN flake. Dotted line shows the thin region of interest. (c) hBN flake transferred on to a set of  bottom electrodes (d) Tunnel device with the horizontal top electrode forming multiple tunnel junctions, one of which is marked as a box. Scale bar is 10 $\mu$m.} 
\end{figure}

In a conventional metal-insulator-metal (M-I-M) junction, the tunnel current-voltage characteristic (\iv) is linear at low biases, with the tunneling resistance inversely proportional to sample area and exponentially dependent on the barrier thickness $d$. Our studies show that the \iv\ with hBN as the barrier is distinct from such a simple behavior in various respects. Thinner barrier devices show a finite linear tunnel current at low biases and a roughly exponential dependence of the low bias resistance with $d$, complying with standard quantum mechanical tunneling, in agreement with previous reports\cite{britnell12,fallahazad14}. However, in relatively thicker barrier devices, we find signatures of Coulomb blockade and single electron tunneling events. In contrast to the conventional systems that show such effects, namely metallic islands and semiconductor quantum dots at low temperatures~\cite{kouwenhoven97}, strikingly similar features are seen here in a presumably large area planar M-I-M junction. 

We have primarily investigated tunnel junctions comprising Cr/Au-hBN-Cr/Au (see Fig. 1a) with 2-6 atomic layers of hBN forming the insulating layer, beyond which tunnel current was unobservably low. The device fabrication begins by mechanically exfoliating hBN flakes on a Si/SiO$_2$ wafer. Figure 1b shows a typical hBN flake. These are further characterized by a combination of optical microscopy, atomic force microscopy and Raman spectroscopy to determine the number of layers. We note that employing a Nomarski interference contrast microscope helps in identifying thinner layers of hBN, which have lower optical contrast in comparison to graphene. The hBN flake is then transferred to another Si/SiO$_2$ wafer previously patterned with narrow Cr/Au electrodes, shown as vertical gold lines in Fig. 1c. We have used the dry transfer method developed by Wang et. al\cite{wang13} for all our devices. The top Cr/Au electrode (horizontal gold pad in Fig. 1d) is then lithographically patterned. In all the devices, there are several bottom electrodes and one top electrode, thus forming multiple tunnel junctions (one out of four tunnel junctions in Fig. 1d is indicated by the box) each of which can be measured in a four-terminal geometry. Typical junctions vary from 1.4 x 0.5 $\mu$m$^2$ to 4 x 2 $\mu$m$^2$ in area.  
The tunneling current $I$ and differential conductance $dI/dV$ in response to voltage excitation ($V=V_{dc}+V_{ac}$, with $V_{ac}=0.5$ mV at 13 Hz) were recorded simultaneously. Three different batches of hBN crystals were used for the present study and no qualitative differences were seen. In addition to junctions with Cr/Au electrodes, we have also examined graphite-hBN-graphite devices.  All of the data were taken at $T=4.2$ K.

\begin{figure}
\centering
\includegraphics[scale=0.47]{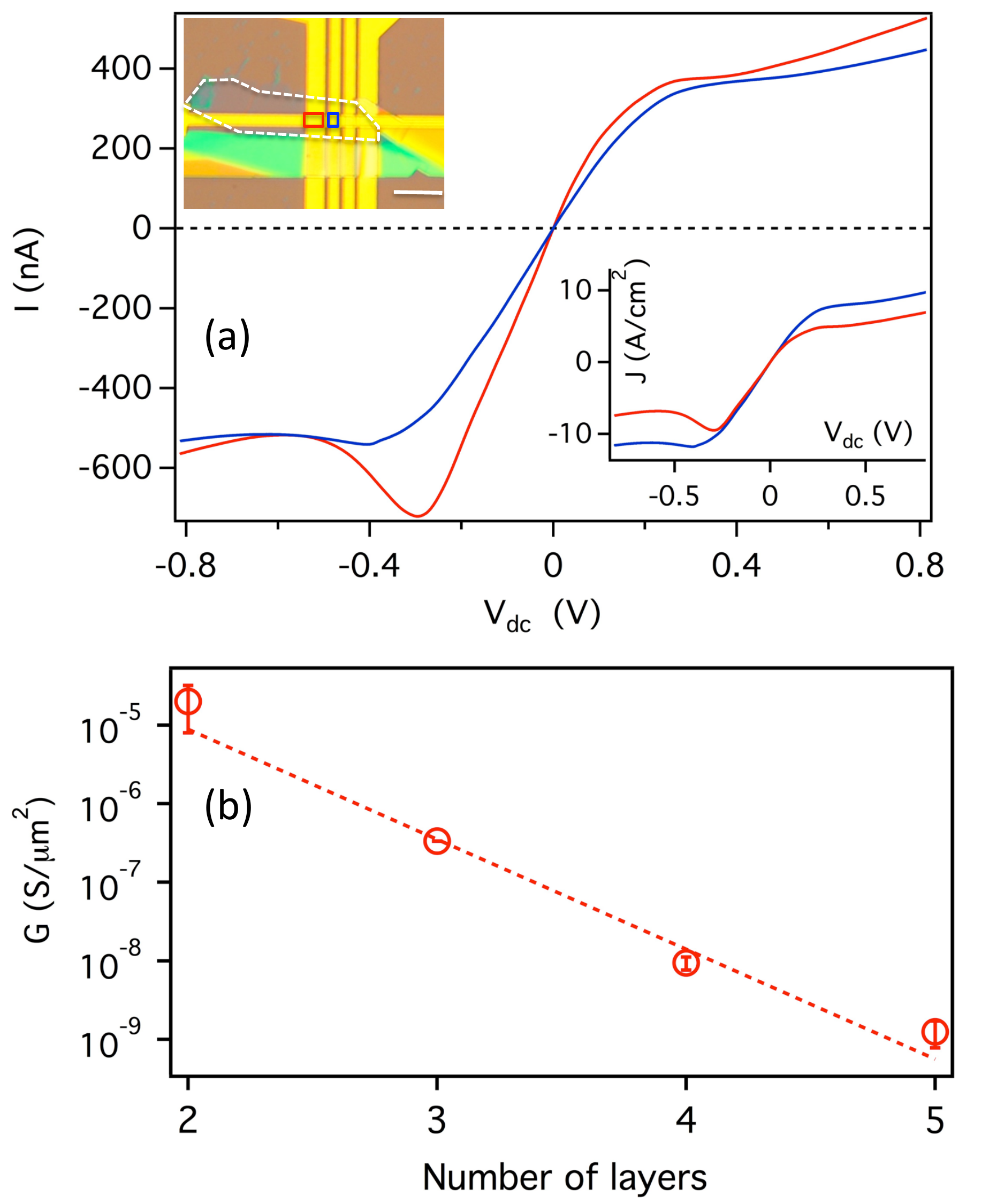}
\caption{(a) Tunnel current ($I$) as a function of dc bias ($V_{dc}$) for a typical thin hBN (dotted line) device with $\approx$ 3 layers. Two tunnel junctions are shown in the inset optical micrograph, with areas 3.8 x 2 $\mu$m$^2$ (red) and 2.3 x 2 $\mu$m$^2$ (blue). Scale bar is 10 $\mu$m. The inset graph shows the tunnel current density ($J$) for the two junctions as a function of bias.  (b) Conductance per unit area $G$ as a function of number of layers. Dotted line indicates an exponential fit.} 
\end{figure}

We first focus on thinner hBN devices with $<$5 atomic layers (each layer being $\sim 0.34$ nm thick\cite{lee11}.) Ten junctions (from four separate hBN flakes) with varying $d$ and area $A$ were tested. The \iv\ curves show an ohmic dependence at dc bias $V_{dc}<\pm$ 100 mV in all the junctions tested. (At higher biases, the \iv\ characteristics show substantial non-linearity.  We will return to this later.)  Figure 2a presents a optical-micrograph of a three atomic layer hBN device with junctions of varying area, along with \iv\ data from two of the junctions.  At low bias voltages both junctions display a linear \iv. The lower inset to the figure shows that the current {\it density} $J=I/A$ for the two junctions is the same in the linear regime $|V_{dc}| \lesssim 100$ mV, as expected for a simple tunnel junction.  

At low biases with negligible barrier deformation, the linear current density for a M-I-M barrier\cite{Simmons63} is,
\begin{equation}
J\propto\exp\Big({\frac{-2d\sqrt{2m^*\phi_B}}{\hbar}}\Big)V,
\end{equation}
where $\phi_B$ is the tunnel barrier height and $m*$ is the effective mass.  Figure 2b reveals that the low-bias tunneling conductance per unit area $G=J/V$ in our thinner barrier devices exhibits this basic exponential dependence.  Not surprisingly, small variations in $G$ among the various junctions with same $d$ are observed, shown by error bars in Fig.2b.  These variations might be due to undetected variations in the thickness $d$ of the hBN layer or lack of uniform contact with the gold electrodes.  An estimate to the factor $m^*\phi_B$ can be obtained by fitting a simple exponential to the  $G$ {\it vs.} $d$ data. Assuming the effective mass to be 0.26$m_0$ for the conduction band edge in hBN~\cite{xu91}, we get $\phi_B=3.3$ eV.  This seems reasonable given the presently unknown alignment of the $\sim 5.9$ eV hBN band gap relative to the Fermi level in the electrodes and the appropriate effective mass for an electron tunneling far from the hBN conduction and valence band edges.  

\begin{figure}
\centering
\includegraphics[scale=0.45]{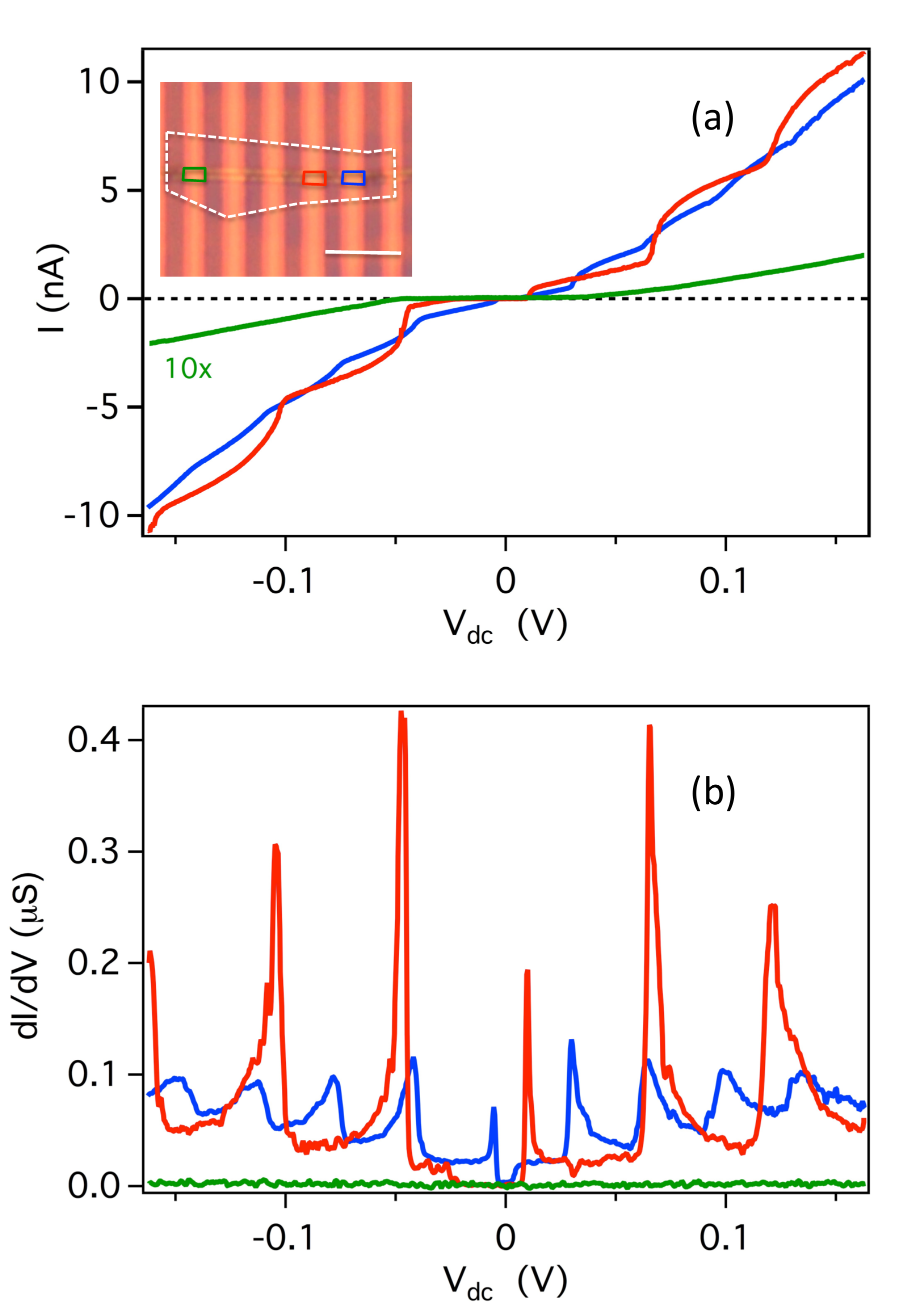}
\caption{(a) Tunnel current ($I$) and (b) differential conductance ($dI/dV$) as a function of dc bias ($V_{dc}$) for a typical thick hBN device with $\approx$ 6 layers. Green, blue and red denote three junctions with area 1.4 x 0.5 $\mu$m$^2$ as shown in the inset. Scale bar is 5 $\mu$m.} 
\end{figure}

We now turn to tunneling transport in samples with thicker hBN barrier layers ($\geq$ 6 layers). Figure 3a shows a typical device with five tunnel junctions of area 1.4 x 0.5 $\mu$m$^2$. Here, we find large discrepancy in the quantitative values of the tunnel current density between the different junctions that are geometrically identical. While most of the junctions had a clear suppressed tunneling region around zero bias (green), some of the junctions also showed discrete steps in the \iv\ (red and blue). Figure 3b shows $dI/dV$ as a function of $V_{dc}$. Consistent with the steps in \iv, periodic oscillations are seen with an energy spacing $\approx 60$ meV. This zero-bias suppression and staircase pattern in \iv\ bear strong resemblance to Coulomb blockade features seen in metallic or semiconducting dots~\cite{kouwenhoven97}. While the suppression of current around zero bias is a universal feature of Coulomb blockade, the Coulomb staircase constitutes a rather special case where the tunnel barriers that isolate the island from the leads are asymmetric (with unequal tunneling resistances and/or capacitances)~\cite{averin91, hanna91}. In such a scenario, an extra electron dwells in the island as the transmittance through one of the barriers is suppressed. Coulomb staircase patterns are observed with voltage spacing given by $e/C$, where $C=C_1+C_2+C_{s}$, $C_1$ and $C_2$ being the capacitances to the source and drain leads and $C_{s}$ is the self capacitance of the island. The tunneling characteristics showed very little dependence on temperature or magnetic field, with the Coulomb blockade features disappearing above about 77 K. 

Sharp features in $dI/dV$ arising from various inelastic processes involving phonons have long been used as a sensitive tool to probe the phonon spectrum of the insulator, the electrodes or molecules trapped at the interface~\cite{chynoweth62}. 
However, the periodicity that we find in the structures, along with the strong blockaded region around zero bias suggest that these are most likely signatures of single electron charging events. The charging energy  $E_c=e^2/2C$ is typically between $20-100$ meV in most samples, giving a capacitance $\approx 0.8-4$ aF, which indicates a very small region as the main contributor to the transport, even though the tunnel junction itself is of the order of microns. Ignoring the capacitance to the leads for now, the self capacitance of a disc of radius $r$ is given by $C_{s}=8\epsilon_0 \epsilon_r r$, which for a capacitance of $1$ aF and $\epsilon_r=4$, would give a radius of $3.5$ nm. The question to ask then, is what component here acts as a dot or an island separated from the metal electrodes? Interestingly, one possible candidate that complies with the small size of the island are the intrinsic defects in the hBN layer. hBN crystals have been known to contain carbon and oxygen impurities in the order of 10$^{17}$/cm$^3$, which have been experimentally determined by secondary ion mass spectroscopy and cathode luminescence~\cite{taniguchi07}. This would correspond to about 100 such defects in a 1x1 $\mu$m$^2$ area for a 1 nm thin sample of hBN. In addition, transmission electron microscopy (TEM) studies have revealed vacancies, interstitial defects and ionized centers in hBN samples~\cite{alem11,cretu14,jin09}. Recent reports of photo-induced doping in graphene on hBN substrates~\cite{ju14} and scanning tunneling microscopy studies of defects in exfoliated hBN~\cite{wong14} substantiates this picture. Importantly, defects and voids in hBN can even lead to interlayer bonding in contrast to few layer graphene sheets~\cite{Rasool15}. These suggest that defects can play a major role in tunneling transport as well, forming small isolated islands in the insulating matrix acting like quantum dots, where adding an extra electron costs energy. We thus have two scenarios: (1) conventional direct tunneling that scales with area and depends exponentially on the barrier thickness and (2) tunneling mediated by defects. The former is more pronounced in thinner samples, while the latter contributes to the sharp features in $dI/dV$ in thicker samples where direct tunneling is suppressed. Note that the capacitance to the electrodes is a relevant quantity here, as the hBN barrier is very thin with a thickness of $\approx 1.8-2$ nm. The capacitance to the leads for a disk of radius 3 nm is given by $C_1=C_2\sim A\epsilon_0 \epsilon_r/d=1$ aF, for $d=1$ nm while, the self capacitance $C_s$ = 0.8 aF.   Indeed, the metal electrodes in our layered tunnel junctions very effectively screen the hBN from externally applied electric fields.  This has the unfortunate consequence that the ``Coulomb diamonds'' which are characteristic of single electron tunneling events, cannot be observed in our devices.  

\begin{figure}
\centering
\includegraphics[scale=0.7]{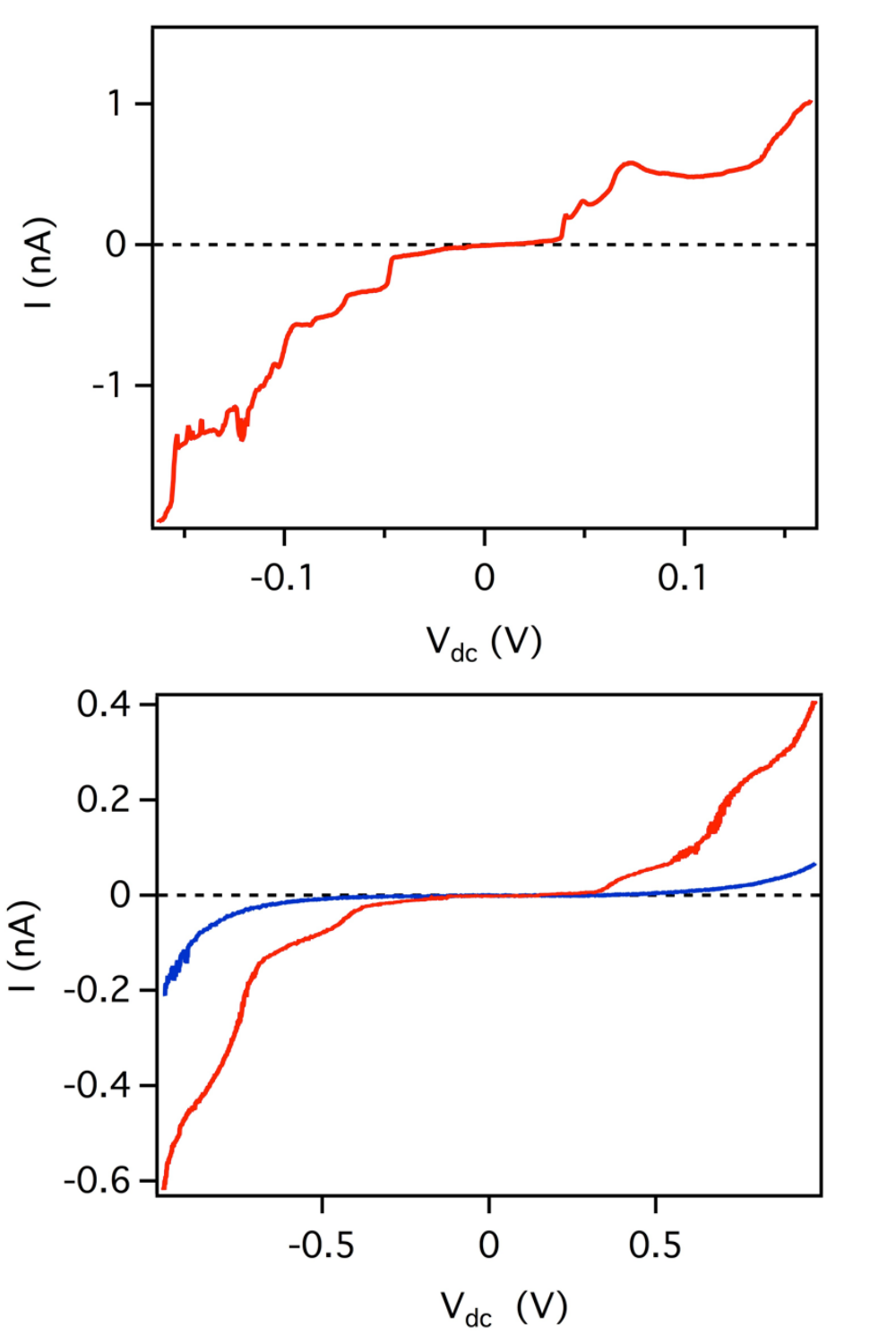}
\caption{(a) Tunneling \iv\ for a typical graphite-hBN-graphite junction with an area 2 x 1.5 $\mu$m$^2$. Two other junctions showed similar characteristics. (b) \iv\ for a thick hBN barrier tunnel junction with Cr/Au electrodes, before (red) and after (blue) annealing in Ar-H$_{2}$ environment. Similar results were obtained for three more junctions in two sets of devices.} 
\end{figure}

In order to eliminate any possible influence of the electrode metal or the fabrication process, we tested a graphite-hBN-graphite device consisting of three junctions with $\approx 6$ atomic layers of hBN. Bottom graphite electrodes were fabricated by mechanical exfoliation followed by e-beam lithography and oxygen plasma etch to create multiple electrodes. The hBN flake was deposited on  the bottom graphite electrodes, followed by the transfer and fabrication of a top graphite electrode. Graphite offers an atomically flat electrode surface for the tunnel junction. As can be seen from Fig. 4a, we find very similar characteristics even in this case, with a blockaded region around zero bias and step-like characteristics in the \iv. Thus, a bad contact with the Au electrodes or the possibility of a Au dot in the hBN layers due to the fabrication procedure can be safely dismissed.

\begin{figure}
\centering
\includegraphics[scale=0.31]{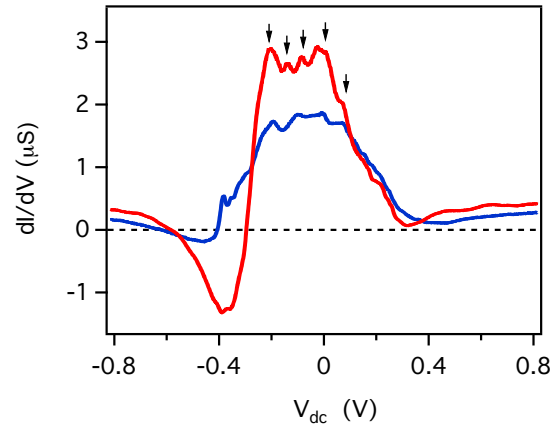}
   \vspace{-0.3in}
\caption{Differential conductance $dI/dV$ {\it vs.} $V_{dc}$ for the thin barrier device shown in Fig. 2a. Red and blue denote the two junctions shown in the inset to Fig. 2a.} 
\end{figure}

In addition to probing the signatures of defects in charge transport, we also find that we can manipulate them by external means. To achieve this, we annealed some of the devices that showed significant Coulomb blockade features in an Ar:H$_2$ environment at 350$^0$C for 3-4 hours. Figure 4b shows the results for one such junction; similar results were obtained from other junctions as well. As the figure reveals, the pristine sample, for which the hBN layer was sufficiently thick that no direct tunneling was expected or observed, showed pronounced Coulomb blockade and staircase features, with $E_c \approx$ 100 meV.  After annealing all signatures of single electron charging effects are absent.  Only at relatively large bias, $|V_{dc}|\sim 0.6$ V, does a significant tunneling current gradually appear. In graphene and other layered 2D materials, annealing in an atmosphere of Ar:H$_2$ or Ar:O$_2$ is expected to remove leftover polymers and other residues from device fabrication and improve adhesion to the substrate\cite{Lin12}. Annealing the exfoliated hBN before transferring graphene on it is often considered an important step in achieving high mobilities for graphene devices\cite{Dean10}. However, the microscopic mechanisms behind such drastic improvements with annealing are still not fully clear. In our devices, annealing seems to eliminate the defects in the active tunnel junction area. In bulk hBN, vacancy migration is expected to occur around 500$^0$C\cite{Zobelli07}. Recent TEM studies on atomic layers of hBN have revealed grain boundaries, square-octagon (4$\vert$8) defects, pentagon-heptagon defects (5$\vert$7) in addition to vacancies and interstitials~\cite{Gibb13,cretu14}. While experiments have revealed that $4\vert$8 defects are mobile at about 800$^0$C under e-beam irradiation\cite{cretu14}, $5\vert$7 defects are expected to be move to the boundary and stop or vanish\cite{Wang15}. Whether annealing cures some of the defects or mobilizes them will need a microscopic analysis using techniques like TEM and is beyond the scope of this work. Nevertheless, this indicates that the features have an intrinsic origin and can also be tuned via simple annealing steps. In addition, these are the first signatures of a direct influence of hBN on electrical transport in Van der Waals heterostructures.

We now return to the thin barrier devices at large bias voltages. Beyond the ohmic low bias region, the \iv\ curves showed a pronounced non-monotonic region, in many cases resembling an NDR feature as seen in Fig. 2a. Figure 5 shows the differential conductance $dI/dV$ as a function of $V_{dc}$ for the two junctions discussed in Fig. 2a, where the non-ohmic behavior can be seen beyond $V_{dc}\approx 0.2-0.3$ V. While the red curve shows a clear NDR region in the negative voltage side, the effect is less pronounced in the positive side and for the blue curve. The voltage range probed is believed to be below the Fowler-Nordheim regime, where the barrier is essentially triangular, due to a very high applied bias\cite{lee11}. In recent studies, NDR signatures were observed in single and bilayer graphene based heterostructures, which were attributed to resonant tunneling via momentum conservation when the energy bands of the top and bottom graphene layers were aligned~\cite{mischchenko14,fallahazad14}. However, we find similar \iv\ curves in simple M-I-M junctions here, albeit with peak to valley ratios lower than in graphene based devices. Interestingly, the only common feature in all these devices is the hBN barrier layer. The origin of these non-ohmic features remains ambiguous. Finally, we also observe several weak peaks at regular intervals in the $dI/dV$ (marked with arrows), in many of the samples.  These may be due to defect-mediated single electron charging events adding to the much larger direct tunnel current.  

In conclusion, we have reported tunneling transport experiments on simple tunnel junctions consisting of metal (or graphite) electrodes separated by hexagonal boron nitride barrier layers.  For thin hBN layers, direct tunneling, with reasonable area and barrier width scaling is observed at low biases, while unexplained non-linear effects are seen at higher biases.  In junctions with thicker hBN barriers, direct tunneling is negligible but strong evidence for defect-mediated single electron tunneling processes is found.  The defect states, and hence the tunneling transport in these thicker barrier devices, can be altered through a simple annealing process. 

\begin{acknowledgements}
We thank J. Petta, J. Velasco Jr. and D. Wong for useful discussions. Special thanks to G. Rossman for the use of his Raman spectroscopy facility. Atomic force microscopy was done at the Molecular Materials Research Center of the Beckman Institute at the California Institute of Technology. This work was supported by the Institute for Quantum Information and Matter, an NSF Physics Frontiers Center with support of the Gordon and Betty Moore Foundation through Grant No. GBMF1250.
\end{acknowledgements}

\end{document}